\documentclass[]{spie}
\usepackage[]{graphicx}

\title{Peter Scheglov --- pioneer of site testing in the Central Asia}
\author{ A.Tokovinin \supit{a} \& V.Kornilov\supit{b}
\skiplinehalf
\supit{a}Cerro Tololo Inrteramerican Observatory, Universitetsky pr-t, 13, La Serena, Chile\\
\supit{b}Sternberg Astronomical Institute, Universitetsky pr-t, 13, Moscow, Russia
}
\authorinfo{Send correspondence to E-mail: victor@sai.msu.ru}
\begin{document}
\maketitle
\begin{abstract}
The multi-faceted  contributions of Dr. Peter  Scheglov (1932-2002) in
the  area of  site testing  are briefly  reviewed.  He  discovered and
studied  astronomical  sites  in   the  Central  Asia,  developed  new
site-testing  instruments, promoted new  methods and  techniques among
his   colleagues   and  teached   new   generation  of   observational
astronomers.
\end{abstract}


\section{Chronology}

\begin{figure}[h]
\begin{tabular}{cc}
\parbox[b]{9.5cm}{ 
\begin{itemize}\itemsep=-3pt
\item[1932 --] On September 4, 1932, Petr (Peter) Vladimirovich Scheglov is born in Tashkent (Uzbekistan)
\item[1954 --] P.S. graduated from the Moscow University, chair of Astrophysics
\item[1957 --] PhD in astronomy (adviser -- I.S.~Shklovsky). P.S. starts working at the Sternberg Astronomical Institute (Moscow).
\item[1966 --] P.S. meets with Jurgen Stock at the IAU General Assembly in Prague, sparkling his interest in site-testing
\item[1967-70 --] Development of the double-beam instrument (DBI) and first site-testing missions to Maidanak, Sanglok, Alma-Ata and Crimea.
\item[1970 --] Dr.Sci dissertation
\item[1974 --] Sternberg Institude decides to build its observatory at Maidanak.
\item[1975 --] Another mission, choice of the observatory location on the West summit of Maidanak.
\item[1976-80 --] Development of the photoelectric seeing monitor (FEP)
\item[1980s --] Seeing measurements at various sites. Evaluation of the ground-layer turbulence with micro-thermal sensors and Sodar.
\item[1980 --] P.S. publishes his book ``Problems of optical astronomy''
\item[1990 --] Comprehensive study of turbulence at Maidanak
\item[2002 --] On December 2, 2002 Petr Vladimirovich Scheglov passed away
\end{itemize}
} &
\includegraphics[height=10cm]{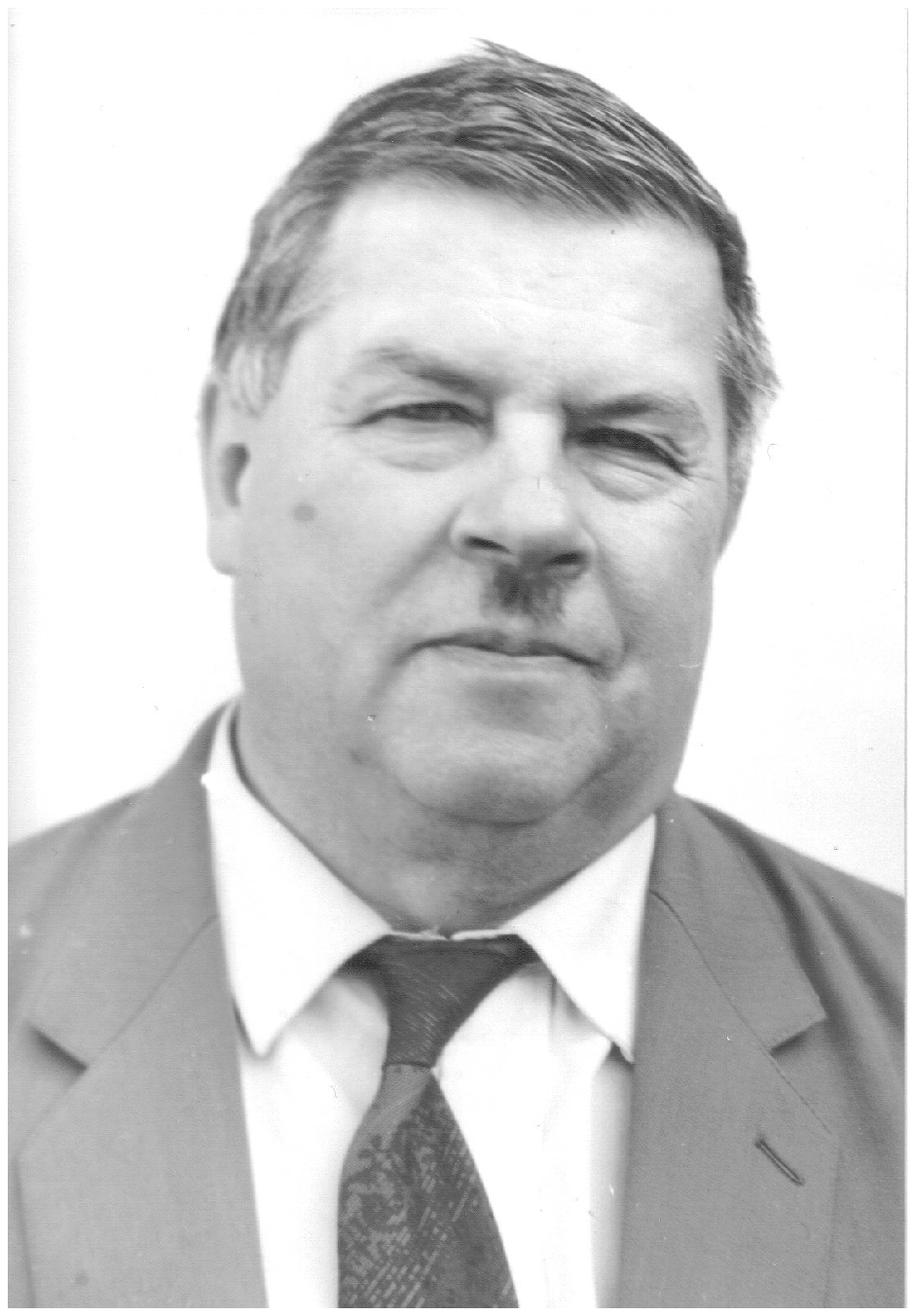}\\
\end{tabular}
\end{figure}

\section{Promote astronomy in the Central Asia. }

\begin{figure}[t]
\begin{center}
\includegraphics[height=7.8cm]{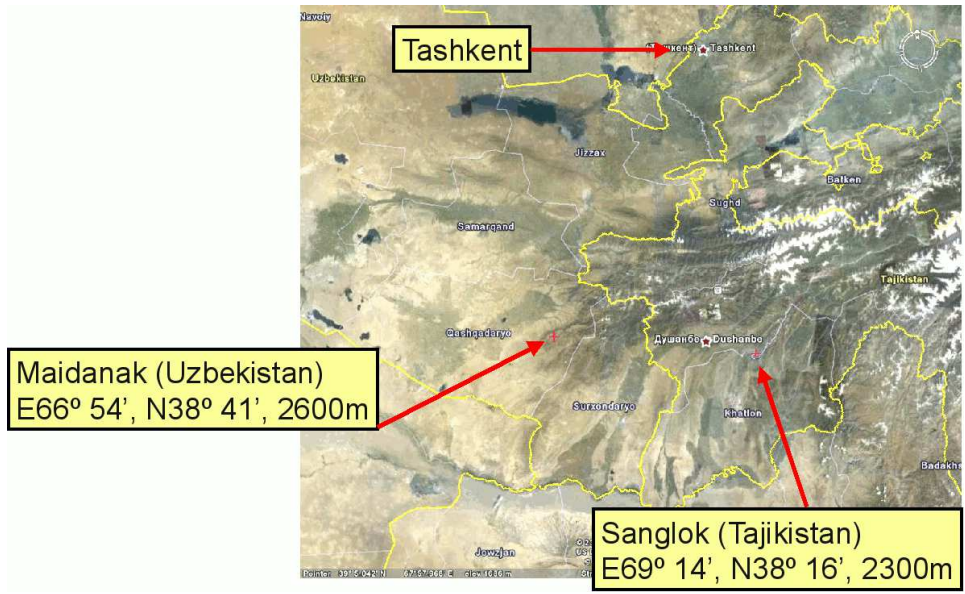}
\end{center}
\caption[example] {Central Asian sites where  P.~Scheglov worked at different times. He studied carefully both marked places \label{fig:2} }
\end{figure}

P.~Scheglov   has  identified   several  Central   Asian   sites  (see
Fig.~\ref{fig:2})  in  collaboration  with meteorologists.   The  most
known  are   Maidanak  (nowadays   an  observatory  of   the  Tashkent
Astronomical  Institute) and  Sanglok  (the observatory  of the  Tajik
Astronomical  Institute). These  sties  are characterized  by a  large
fraction of clear  sky (in excess of 2000 night  hours per year), good
seeing, and calm conditions in the upper atmosphere. This latter fact,
due  to the  low jet-stream  activity in  the region,  is particularly
valuable   for   high-resolution   techniques.    Scheglov   organized
site-testing expeditions, helped  the development of new observatories
and education of astronomy in the Soviet Central Asia and Mongolia.

\begin{figure}[b]
\begin{center}
\includegraphics[height=6cm]{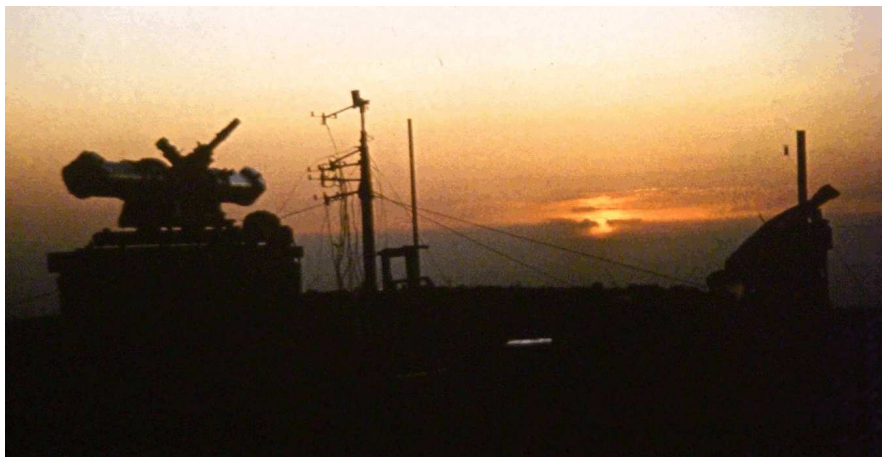}
\end{center}
\caption[example]  {Site-testing  expedition  at  the West  Summit  of
  Maidanak (1976)  with silouettes of  the DBI (left),  the telescopic
  micro-thermal mast (center) and the FEP (right).
\label{fig:4} }
\end{figure}

\section{Develop site-testing methods and instruments. }

P.~Scheglov  developed site-testing methods  and instruments  and used
them at numerous observatories.  He started by copying the double-beam
instrument  of J.~Stock (with  his PhD  student S.~Novikov),  lead the
development  of the  first photoelectric  image-motion monitor  with a
servo   system  (1975),   later  developed   and  actively   used  the
photoelectric seeing monitor  (FEP) with a fixed knife  working on the
Polar Star. At the same  time, he established close collaboration with
the Institute  of Atmospheric  Physics (IAP), at  that time  the world
leaders in turbulence theory and measurement.  Sensitive micro-thermal
system was created by his PhD student A.~Guryanov.  Scheglov organized
the participation  of the  IAP SODAR team  in the  characterization of
Sanglok, Maidanak, and of the Russian 6-m telescope in Caucasus.

\begin{figure}
\begin{center}
\begin{tabular}{cc}
\includegraphics[height=7.8cm]{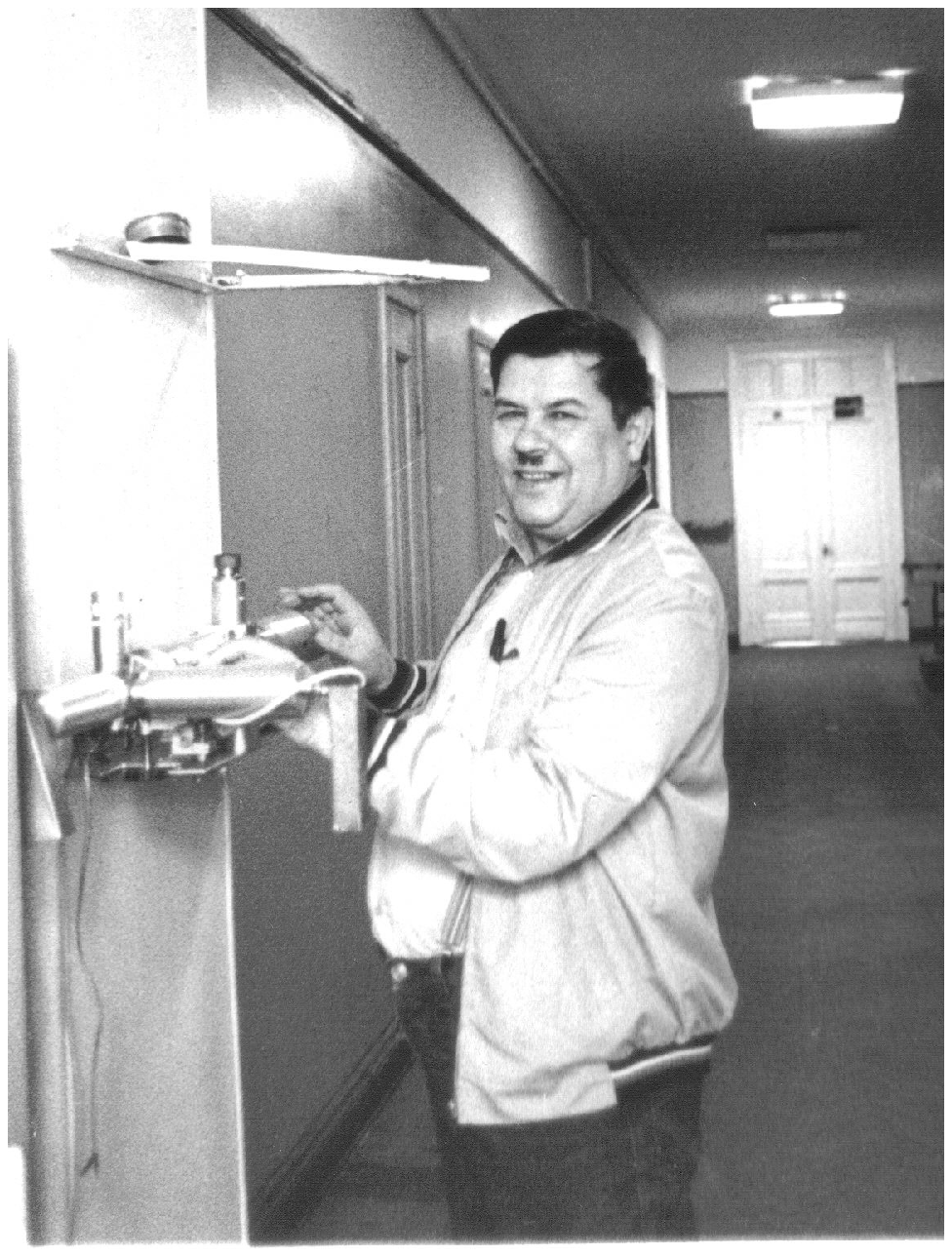} &
\includegraphics[height=7.8cm]{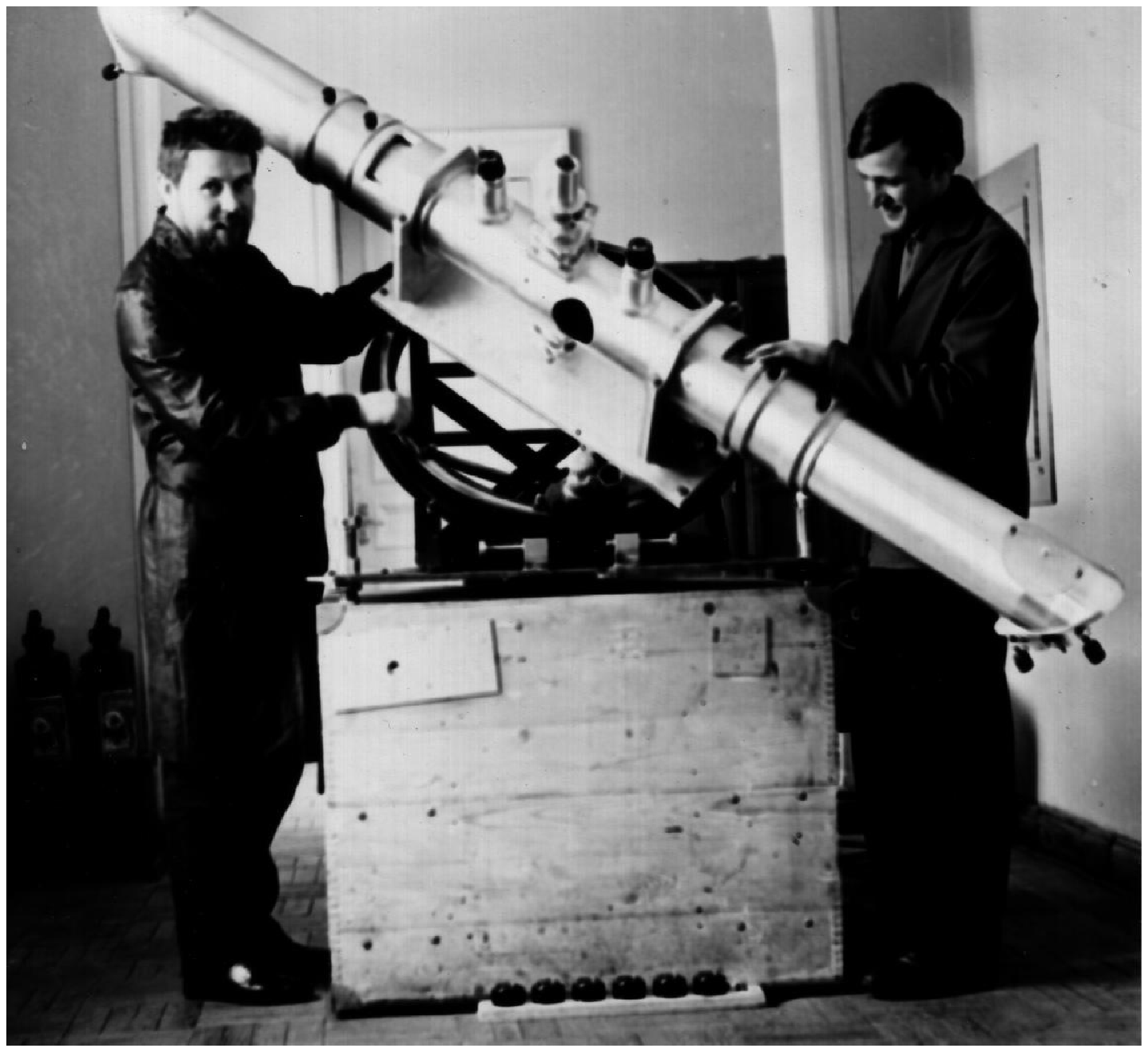} \\
\end{tabular}
\end{center}
\caption[example]  {{\it  Left: }  Calibration  of the  Photo-Electric
  Seeing Monitor.  {\it Right: } P.~Scheglov and S.~Novikov with the
  double-beam instrument (DBI), 1960s \label{fig:3} }
\end{figure}

\begin{figure}[b]
\begin{center}
\begin{tabular}{cc}
\includegraphics[height=5.8cm]{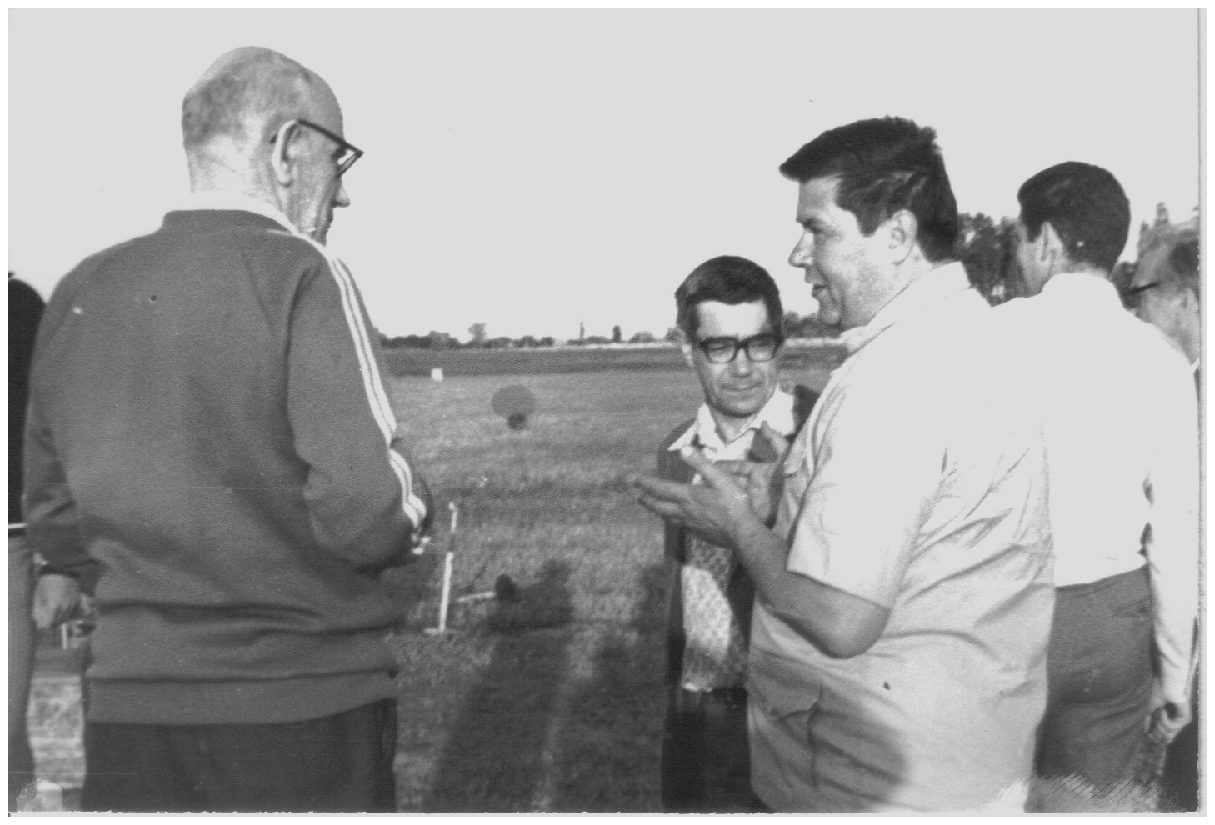}&
\includegraphics[height=5.8cm]{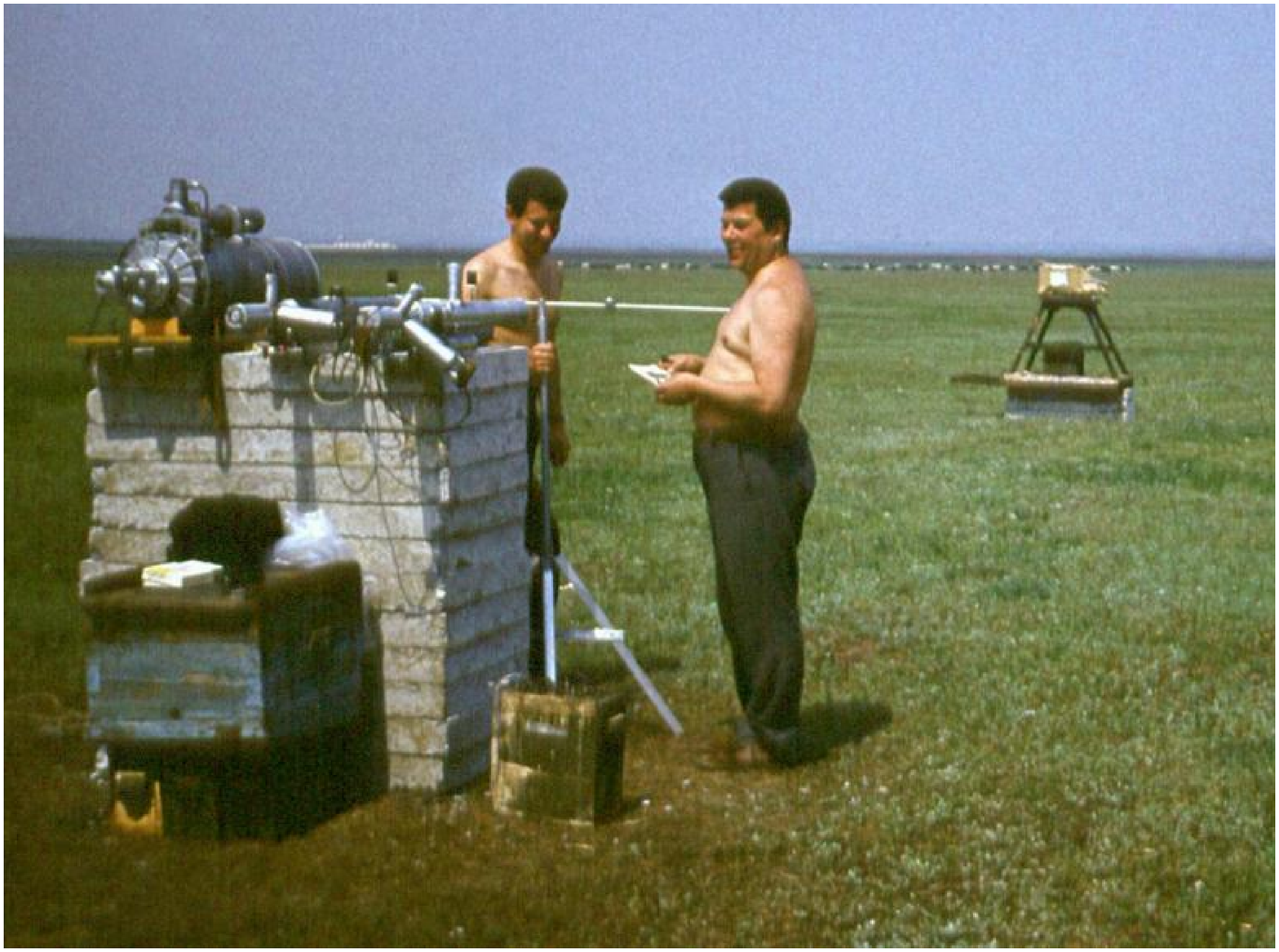}\\
\end{tabular}
\end{center}
\caption[example] { Calibration of optical and micro-thermal equipment
  in   Tsimliansk  (May   1982).  Left:   A.M.~Obukhov,  V.N.~Karpinsky,
  P.V.~Scheglov,  A.E.~Guryanov.  Right:  Guryanov and  Scheglov prepare
  experiments on the horizontal path.\label{fig:6} }
\end{figure}

\section{Towards high angular resolution in astronomy. }

P.~Scheglov  worked on increasing  angular resolution  of ground-based
telescopes, realizing that the sensitivity to faint objects depends on
the resolution as much as on the telescope diameter. He considered
this to be a huge and poorly used resource of ground-based
astronomy. He approached the problem from multiple sides. 

\begin{itemize}
\item
Select astronomical sites with best natural seeing.

\item
Study the man-made seeing in domes and telescopes, determine optimum
parameters of astronomical domes (height, ventillation regime).

\item
Interact with  the optical industry  to help build telescopes  of high
optical quality.  P.S.  had a  strong interest in optical  testing and
telescope technology.

\item
Develop electronic  and photographic detectors to  approach the photon
sensitivity limits. P.S. was a  pioneer of using image intensifiers in
astronomy.

\item
Increase the resolution by post-processing (image sharpening) and
interferometry. P.S. was among the first to recognize and advocate
the potential of speckle interferometry and long-baseline
interferometers. 

\end{itemize}

P.S.   tried to  actively promote  these ideas  by presenting  them at
various  colloquia  and meetings,  publishing  review papers,  Russian
translations  of  textbooks and  proceedings  on  detectors and  large
telescopes of  the future and, finally,  writing his own  books on new
astronomical techniques.

\begin{figure}
\begin{center}
\includegraphics[height=7cm]{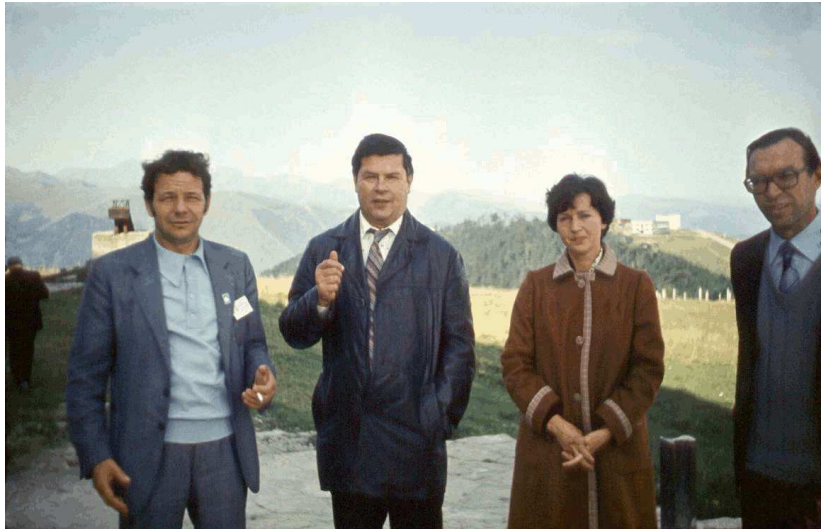}
\end{center}
\caption[example]   {At  the   IAU  Colloquium   67   on  astronomical
  instruments   (September  1981,  at the 6-m telescope):   V.N.~Dudinov,  P.V.~Scheglov,
  V.S.~Tsvetkova, V.F.~Esipov \label{fig:7} }
\end{figure}

\section{Prepare next generation of astronomers. }

P.~Scheglov  teached   at  the  Moscow  University.    His  course  on
experimental astronomy  largely contributed  to the education  of many
currently  active researchers.   He  was the  PhD  advisor of  several
graduate  students working on  site testing:  S.~Novikov, A.~Guryanov,
V.~Kornilov,  A.~Tokovinin, A.~Kutyrev,  Yu.~Khan.  He  has influenced
the whole  generation of  astronomers in the  former republics  of the
Soviet Union.

\section{Selected references}

P.V. Scheglov (Sheglov, Shcheglov) published about 50 papers on site testing, some are listed below.

Novikov, S.B., Sheglov, P.V. 1968, Preliminaty results of double-beam site testing at Mt.Sanglock.  Astr. Tsirk., No. 491, 3

Efremov, Iu.N., Novikov, S.B., Shcheglov, P.V. Prospects for development of ground-based optical astronomy. 1975. Sov. Phys. Uspekhi, 18, 151

Shcheglov, P.V. On the use of acoustic methods for studying temperature and wind fields near astronomical instruments. 1976, Astr. Tsirk. No. 900, 3

Beslik, A.I. et al.  Simultaneous seeing measurements near Mt. Maidanak with the double-beam telescope and a photoelectric device. 1977, Astr. Tsirk., No. 955, 3

Scheglov P.V. Astroclimatic studies in the Soviet part of the Central Asia. Proc. Conf. on Astroclimate, Abastumani, 23-26 Nov. 1981.

Guryanov et al. A complex study of optically active turbulence above two mountain observatories. 1988, Astron. Zh., 65, 637

Guryanov et al.  The contribution of the lower atmospheric layers to the seeing at some mountain observatories. 1992, A\&A, 262, 37

Shcheglov, P.V. On some features of atmospheric circulation favorable to good seeing. 1998, Astron. Tsirk., No. 1557.

\end{document}